\documentstyle[twoside,fleqn,npb,epsfig]{article}

\title{Transversity distributions and Drell-Yan spin asymmetries}

\author{M. Miyama
        \address{Department of Physics, Saga University,
                 Saga 840-8502, Japan}
        \thanks{Research Fellow of the Japan Society for the 
                Promotion of Science. 
                E-mail: miyama@cc.saga-u.ac.jp.}}

\begin{document}


\pagestyle{empty}

\begin{flushleft}
\Large
{SAGA-HE-148-1999
\hfill May 31, 1999}  \\
\end{flushleft}

\vspace{1.6cm}
\begin{center}
 
\LARGE{{\bf Transversity distributions}} \\

\vspace{0.2cm}
\LARGE{{\bf   and Drell-Yan spin asymmetries}} \\

\vspace{1.1cm}
\LARGE
{M. Miyama $^*$} \\
 
\vspace{0.3cm}
\LARGE
{Department of Physics} \\
 
\LARGE
{Saga University} \\
 
\LARGE
{Saga 840-8502, Japan} \\

\vspace{1.0cm}
 
\LARGE
{Talk given at the 7th International Workshop} \\

\vspace{0.1cm}

{on Deep Inelastic Scattering and QCD} \\

\vspace{0.1cm}

{DESY Zeuthen, Germany, April 19 -- 23, 1999} \\

\vspace{0.05cm}
{(talk on April 22, 1999) }  \\
 
\end{center}
 
\vspace{0.7cm}
\vfill
 
\noindent
{\rule{6.0cm}{0.1mm}} \\
 
\vspace{-0.3cm}
\normalsize
\noindent
{* Email: miyama@cc.saga-u.ac.jp. 
Information on their research is available at } \\

\vspace{-0.44cm}
\noindent
{\ \ \  http://www2.cc.saga-u.ac.jp/saga-u/riko/physics/quantum1/
structure.html.}  \\

\vspace{+0.1cm}
\hfill
{\large to be published in Nuclear Physics B}

\vfill\eject
\setcounter{page}{1}
\pagestyle{crcplain}

\begin{abstract}

We discuss transversity distributions and Drell-Yan
transverse double spin asymmetries.
First, the antiquark flavor asymmetry 
$\Delta_{_T} \bar u/\Delta_{_T} \bar d$ is discussed by using two 
different descriptions, a meson-cloud model and a Pauli exclusion model.
We find that both calculations produce a significant
$\Delta_{_T} \bar d$ excess over $\Delta_{_T} \bar u$.
Next, we study its effects on the transverse spin asymmetry $A_{_{TT}}$
and on the Drell-Yan proton-deuteron asymmetry 
$\Delta_{_T}\sigma^{pd}/2 \Delta_{_T}\sigma^{pp}$. We find that 
the ratio $\Delta_{_T}\sigma^{pd}/2 \Delta_{_T}\sigma^{pp}$ is 
very useful for investigating the flavor asymmetry effect. 

\end{abstract}

\maketitle

\section{Introduction}

Since the discovery of a serious spin problem, the internal 
spin structure of the nucleon has been a popular topic. 
Using many experimental data on $g_1$, we have a rough idea 
on the longitudinally polarized parton distributions.
On the other hand, the transversity 
distributions $\Delta_{_T} q$ have not been measured at all 
because they cannot be measured in the usual inclusive 
deep inelastic scattering.
They are expected to be measured in the transversely polarized 
Drell-Yan process at RHIC and semi-inclusive process at HERA. 
We should try to understand the properties of $\Delta_{_T} q$ before 
the experimental data are taken.

The $Q^2$ evolution of the transversity distributions has
been already investigated in detail including 
the next-to-leading-order (NLO) effects \cite{HKM}. 
Using these results, we investigate the antiquark flavor asymmetry 
in the transversity distributions \cite{KM}. 
Now, it is well known that light antiquark distributions are not 
flavor symmetric \cite{SK} according to the NMC, NA51, E866, 
and HERMES experimental data. 
In particular, the recent E866 result revealed 
the $x$ dependence of the ratio $\bar d/\bar u$ by measuring
the Drell-Yan proton-deuteron asymmetry.
The antiquark flavor asymmetry in the polarized distributions
is also expected to exist. However, it is not known at this stage 
except for a few theoretical predictions \cite{BS,FS}.
Because the polarized antiquark distributions are measured at RHIC,
it is important to investigate  a possible asymmetric distribution.
In this paper, we calculate the transversity flavor asymmetry
by using the nonperturbative models.
Then, obtained results are used for calculating the transverse
double spin asymmetry $A_{_{TT}}$ at a RHIC energy. 
Furthermore, we discuss the flavor asymmetry effects on the 
Drell-Yan proton-deuteron 
asymmetry $\Delta_{_T}\sigma^{pd}/2\Delta_{_T}\sigma^{pp}$.

\section{Antiquark flavor asymmetry}

The contribution from the perturbative QCD is 
considerably small \cite{HKM} as far as the $Q^2$ evolution 
in the range $Q^2 \geq 1$ GeV$^2$ is concerned. 
Therefore, we expect the antiquark flavor asymmetry 
comes almost from the non-perturbative mechanisms. 
As such mechanisms, a meson-cloud and a Pauli exclusion 
principle models are typical in discussing 
the unpolarized $\bar{u}/\bar{d}$ asymmetry \cite{SK}.
We try to apply these mechanisms to the polarized distributions 
\cite{KM}. Since we have no experimental information about transversity 
distributions, we calculate the flavor asymmetry in the longitudinally
polarized distributions at first. Then, we assume the transversity 
distributions are equal to the longitudinal distributions 
at small $Q^2$ by using the prediction of 
the nonrelativistic quark model.

First, we discuss the meson-cloud model. 
In this model, we calculate the meson-nucleon-baryon (MNB)
process in which the initial nucleon splits into 
a virtual meson and a baryon, then the virtual photon 
from lepton interacts with this meson.
Because the lightest vector meson is the $\rho$ meson, 
we calculate the $\rho$-meson contribution to the flavor asymmetry
in the polarized distributions.
The contributions from this kind of process can be expressed
by the following convolution integral:

\begin{equation}
\Delta\overline{q}(x,Q^2) = \int_{x}^{1} \frac{dy}{y} 
\Delta f_{\rho NB}(y) 
\Delta\overline{q}_\rho (\frac{x}{y}, Q^2) \ ,
\end{equation}

\noindent
where the function $\Delta f_{\rho NB}$ represent the $\rho$-meson 
momentum distribution due to the $\rho$NB process 
and the function $\Delta\overline{q}_\rho$ represent the polarized 
antiquark distribution in the $\rho$ meson. In our analysis, 
nucleon and $\Delta$ are taken into account as a final state baryon 
and all the possible $\rho$NB processes are considered.
Since the dominant contribution comes from the $\rho^+$ meson
which has a valence $\bar d$ quark, the meson-cloud contribute 
$\Delta\bar d$ excess over $\Delta\bar u$.
Note that the $\rho$-meson effects are also studied 
in Ref. \cite{FS} by using the slightly different method 
from ours in the calculation of $\Delta f_{\rho NB}$.

Next, we discuss the Pauli exclusion model.
Because there already exist some studies on this mechanism 
in the polarized case \cite{BS}, we simply use their results.  
According to the SU(6) quark model, 
we have each quark state probability as
$u^\uparrow = 5/3$, $u^\downarrow = 1/3$, $d^\uparrow = 1/3$, 
and $d^\downarrow = 2/3$ in the proton spin-up state.
Since the probability of $u^\uparrow$ ($d^\downarrow$)
is much larger than that of $u^\downarrow$ ($d^\uparrow$),
it is more difficult to create $u^\uparrow$ ($d^\downarrow$) sea 
than $u^\downarrow$ ($d^\uparrow$) sea 
because of the Pauli exclusion principle.
Then, assuming that the exclusion effect is the same as the unpolarized,
$(u_s^\downarrow - u_s^\uparrow)/(u_v^\uparrow - u_v^\downarrow) =
 (d_s - u_s)/(u_v - d_v)$ and a similar equation for 
$d_s^\uparrow - d_s^\downarrow$,
we have $\Delta\overline{u}= -0.13$ and 
$\Delta\overline{d}= +0.05$.
As a result, we find that the flavor asymmetry from this mechanism.

The both models should be valid only at small $Q^2$, so that
the GRSV parametrization at $Q^2$=0.34 GeV$^2$ is chosen 
in our calculation. The obtained results 
for the $\Delta_{_T}\bar u-\Delta_{_T}\bar d$ distribution are shown 
at $Q^2$ = 10 GeV$^2$ in Figure 1.
From this figure, we find that both model predictions
have similar tendency, namely, $\Delta_{_T} \bar d$ excess 
over $\Delta_{_T} \bar u$. However, the meson contributions 
seems to be smaller than those of the exclusion model.
Recently, a flavor asymmetric distribution was proposed by
analysing deep inelastic semi-inclusive data \cite{MY}.
Our distributions are consistent with their result 
because the accuracy of the present semi-inclusive data 
is insufficient to find accurate flavor asymmetry.

\vspace{-0.3cm}
\begin{figure}[htb]
\begin{center}
\epsfig{file=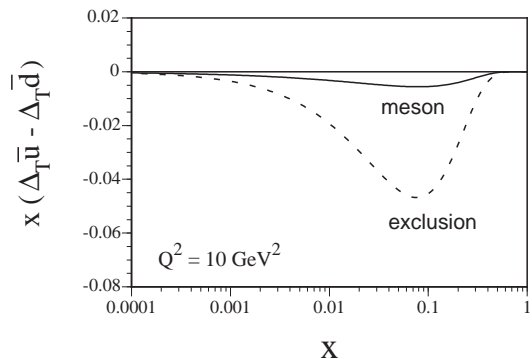,width=7.0cm}
\vspace{-0.5cm}
\caption{Flavor asymmetric distribution.}
\end{center}
\label{fig:1}
\vspace{-0.5cm}
\end{figure}

\section{Transverse double spin asymmetry}

The transversity distributions will be studied by
measuring the Drell-Yan transverse double spin asymmetries 
$A_{_{TT}}$ at RHIC.
In this section, we discuss the flavor asymmetry effects on $A_{_{TT}}$.
We calculate $A_{_{TT}}$ with NLO contributions
at the RHIC energy $\sqrt{s}=200$ GeV by using the flavor symmetric 
and the flavor asymmetric parton distributions.

The results are shown in Figure 2 as the function of dimuon 
mass square. The solid, dashed, and dotted curves represent 
the flavor symmetric, meson-cloud, and Pauli exclusion results, 
respectively.
From this figure, we find that the magnitude of $A_{_{TT}}$ is about 1\%
in the dimuon mass region $100<M_{\mu\mu}^2<500$ GeV$^2$.
Furthermore, we find that the flavor asymmetric results
are considerably different from the flavor symmetric one.
However, the differences are not enough to find
the flavor asymmetry effects because we may change the magnitude of
the transversity distributions so as to agree with the experimental data.
Although the longitudinal flavor asymmetry
$\Delta \bar u/\Delta \bar d$ should be investigated by the $W^\pm$ 
production processes, the transversity asymmetry cannot be measured 
by the $W^\pm$. Therefore, we should think about possible measurements
in order to investigate the $\Delta_{_T}\bar u/\Delta_{_T}\bar d$
asymmetry.

\vspace{-0.3cm}
\begin{figure}[htb]
\begin{center}
\epsfig{file=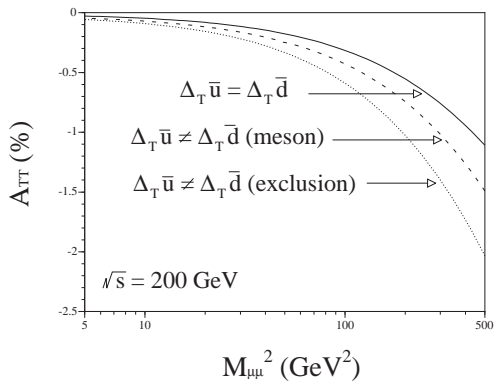,width=6.5cm}
\vspace{-0.5cm}
\caption{Transverse double spin asymmetry.}
\end{center}
\label{fig:2}
\vspace{-0.5cm}
\end{figure}

\section{Drell-Yan proton-deuteron asymmetry}
\
As the alternative candidate for finding the transversity 
flavor asymmetry $\Delta_{_T} \bar u/\Delta_{_T} \bar d$,
we propose to use the polarized proton-deuteron Drell-Yan process 
in combination with the polarized pp process \cite{KM,pd}.
Recently, a general formalism for the polarized pd reactions 
was completed \cite{pd}. From their parton model analysis, 
the Drell-Yan proton-deuteron asymmetry
$\Delta_{_T}\sigma^{pd}/2\Delta_{_T}\sigma^{pp}$ is expressed by
transversity distributions in the proton and the deuteron.
If we neglect the nuclear effects in the deuteron and assume the
isospin symmetry, the Drell-Yan p-d asymmetry 
in the large $x_F$ (=$x_A - x_B$) region
is approximately given by the following equation:

\begin{equation}
\frac{\Delta_{_T}\sigma^{pd}}{2 \Delta_{_T}\sigma^{pp}} \,  
\approx \, \frac{\left[1 + \frac{1}{4} 
\frac{\Delta_{_T}d\left(x_A\right)}{\Delta_{_T}u\left(x_A\right)}\right] 
\left[1 + \frac{\Delta_{_T}\overline{d}\left(x_B\right)}
{\Delta_{_T}\overline{u}\left(x_B\right)}\right]}
{2 \left[1 + \frac{1}{4} 
\frac{\Delta_{_T}d\left(x_A\right)}{\Delta_{_T}u\left(x_A\right)} 
\frac{\Delta_{_T}\overline{d}\left(x_B\right)}
{\Delta_{_T}\overline{u}\left(x_B\right)}\right]} \, ,
\end{equation}

\noindent
where the $x_A$ and $x_B$ represent the momentum fractions in
the hadron A (proton) and the hadron B (deuteron).
From this equation, the ratio clearly becomes one 
if $\Delta_{_T}\bar u$ is equal to $\Delta_{_T}\bar d$.
Therefore, if we find that the data which is different
from one, it suggests the flavor asymmetry.

We calculate the ratio 
$\Delta_{_T}\sigma^{pd}/2\Delta_{_T}\sigma^{pp}$ by using the 
flavor symmetric and the flavor asymmetric distributions \cite{KM}.
As a result, we find that the flavor symmetric result becomes 
almost one in the large $x_F$ region as discussed above 
although the meson-cloud and the Pauli exclusion results 
are significantly different from one.
From this analysis, we conclude that the Drell-Yan p-d asymmetry
is very useful for investigating the light antiquark flavor
asymmetry in the transversity distributions.
We mention that the p-d asymmetry should be also used for
studying the flavor asymmetry in the longitudinally 
polarized distributions in a similar way.
At this stage, there are no project to measure the polarized 
proton-deuteron Drell-Yan process. However, we expect it is 
possible in the future RHIC, FNAL, or HERA experiment.

\vspace{0.5cm}
M.M. was supported by the JSPS Research Fellowships 
for Young Scientists.
This research was partly supported by the Grant-in-Aid 
from the Japanese Ministry of Education, Science, and Culture. 


\end{document}